\documentclass[preprint,12pt,sort&compress]{elsarticle}
\usepackage{amsmath,amssymb}
\DeclareMathOperator{\sech}{sech}
\DeclareMathOperator{\arccosh}{arccosh}

\usepackage{epstopdf}
\usepackage{color}
\usepackage{siunitx}

\begin{document}
\begin{frontmatter}

\title{On solutions of a Boussinesq-type equation with displacement-dependent nonlinearity: a soliton doublet }
\author{Tanel Peets, Kert Tamm, P\"aivo Simson, J\"uri Engelbrecht}
\address{Laboratory of Solid Mechanics, Department of Cybernetics, School of Science\\ Tallinn University of Technology, Akadeemia tee 21, Tallinn 12618, Estonia, \\E-mails: tanelp@ioc.ee, kert@ioc.ee, paivo@sysbio.ioc.ee, je@ioc.ee\\
\flushright To the memory of Alexander Samsonov}

\begin{abstract}
In this paper the permanent profile waves governed by a Boussinesq-type wave equation are analysed. The model involves displacement-type nonlinearities and dispersion terms. Physically such a model equation describes longitudinal waves (density change) in biomembranes which have an internal structure composed by lipid molecules. The possible solutions are constructed and analysed. The phase plane analysis and numerical simulation reveal a novel phenomenon: the possible existence of a soliton doublet. 
\end{abstract}

\end{frontmatter}

\section{Introduction}

Solitons are exceptional examples on complexity of physical world -- waves with a permanent (steady) shape under the influence of several physical effects. As known, solitons may exist in nonlinear dispersive media: fluids, solids, plasma, electrical circuits. The dispersion of media can be caused either by physical properties (for example, existence of microstructures) or geometrical properties (for example, existence of boundaries like in waveguides or free surface like in shallow water). The nonlinearities can also be caused either by physical (stress-strain relations) or by geometrical (large strain) effects. After the pioneering descriptions of solitons by Russell \citep{Russel1844} and Korteweg and de Vries \citep{Korteweg1895}, the contemporary studies started after Zabusky and Kruskal \citep{Zabusky1965} had coined the notion `soliton'. Many fundamental results in solitonics (see, for example, overviews by Newell \citep{Newell1985}, Ablowitz \citep{Ablowitz2011}, etc.) have revealed the properties of solitons. The overview by Maugin \citep{Maugin2011} is focused on the description of solitons in elastic solids.

The governing equations of soliton-type waves are usually of one-wave type \citep{Ablowitz2011}, i.e., first-order evolution equations like the celebrated Korteweg-deVries (KdV) equation. However, the modifications of the classical second-order wave equation derived to include dispersive and nonlinear effects, are also able to describe solitons. Such models are called Boussinesq-type equations \citep{Christov2007}. In wave-guides, the Boussinesq-type models are extensively studied by Samsonov \citep{Samsonov2001} together with the description of experiments demonstrating the existence of solitons in the presence of geometrical dispersion.
Samsonov \citep{Samsonov2001} has stressed that in this case the two-wave governing equation is `a double-dispersion equation' because the derivation of the one-wave evolution equation will not describe properly the dispersive effects. This is also the case of microstructured materials where dispersive effects are caused by microstructural inclusions \citep{Engelbrecht2015a}. 

In this paper, the emergence of solitons in a special biological medium is analysed. This is the case of biomembranes which have an important role in biophysics. The governing equation is of a Boussinesq-type, which like in the case of wave-guides \citep{Samsonov2001} is of the second order with two dispersive terms \citep{Engelbrecht2015} and special nonlinearity \citep{Heimburg2005,Tamm2015}. Such a structure reflects the elastic properties and the microstructure of biomembranes. The previous results \citep{Engelbrecht2017} are enlarged by the detailed analysis of a two-soliton solution which emerges under certain conditions. Such a soliton doublet is a sign of richness in the soliton Zoo.

\section{Mathematical model}

Processes in neurons are complicated phenomena. Classically the function of a nerve has been attributed to an electrical pulse (action potential) propagating along the nerve axon that is understood in terms of the Hodgkin-Huxley model \citep{Hodgkin1952}. In recent decades it has become clear that bioelectricity alone is not sufficient for a complete understanding of the neural function \citep{Mueller2014}. Most prominent nonelectrical effects are a mechanical wave (swelling) \citep{Iwasa1980,Tasaki1988,Gonzalez-Perez2016,Yang2018} and a pressure wave \citep{Terakawa1985} that propagate along the nerve fibre together with the action potential. It is clear that for a complete understanding of nerve function, a model describing all processes in a joint framework is needed \citep{Engelbrecht2018,Engelbrecht2018a}. However, the models describing single waves should be well understood. 

Here we focus our attention to the propagation of mechanical waves in biomembranes which are important constituents or nerve axons and all other living cells \citep{Mueller2014,Blume2018}. These biomembranes are built of phospholipids which in a living tissues  form bilayers with hydrophobic tails directed inward and hydrophilic heads facing outward whereas experiments in labs are often carried out on lipid monolayers \citep{Blume2018}.

 Mechanical waves in lipid bilayers can be modelled by the improved Heimburg-Jackson (HJ) model \citep{Engelbrecht2015,Heimburg2005}:
\begin{equation}
	\label{improvedHJ}
	u_{tt}=\left[(c^2_0+pu+qu^2)u_x\right]_x-h_1u_{xxxx}+h_2u_{xxtt},
\end{equation}
where $u=\Delta\rho_A$ is the longitudinal density change, $c_0$ is the velocity of the unperturbed state, $p$, $q$ are coefficients determined from experiments and $h_1$, $h_2$ are \textit{ad hoc} dispersion coefficients. Here and further indices $x$, $t$ denote partial derivates with respect to space and time, respectively. 

This model (with $h_2=0$) was deduced by Heimburg and Jackson from experimental considerations starting with the regular wave equation and replacing the expression for the effective velocity $c_e$ by $c^2_e=c^2_0+pu+qu^2$ and adding an \emph{ad hoc} dispersive term $h_1u_{xxxx}$ \citep{Heimburg2005}. This model was improved by Engelbrecht et al. by taking inspiration from the rod theories and adding a fourth order mixed derivative term $h_2u_{xxtt}$ \citep{Engelbrecht2015} making the HJ model a `double-dispersion equation' \citep{Samsonov2001} and thus removing instabilities that may arise when only spatial derivatives are present \citep{Maugin1999,Metrikine2002}. Moreover, the mixed derivative term is related to the inertial properties and the spatial derivative term is related to elastic properties of the underlying microstructure (lipid bilayer) \citep{Engelbrecht2015} and both terms arise naturally when proper modelling approach is followed \citep{Engelbrecht2015a,Maurin2016}.

The analytical solution for Eq.~\eqref{improvedHJ} with $h_2=0$ has been derived by Lautrup et al. \citep{Lautrup2011}. They concentrate on the behaviour of lipids in case of $p<0$ and $q>0$ which is a special case when the lipids are above the melting transition \citep{Heimburg2005} resulting in positive amplitude solitary wave solutions. Perez-Camacho et al. \citep{Perez-Camacho2017} have shown that negative amplitude solitary waves can exist also in lipid bilayers and Eq.~\eqref{improvedHJ} provides such solutions when both nonlinear coefficients ($p$ and $q$) are positive. Freist\"uhler and H\"owing \citep{Freistuhler2013} provide a rigorous mathematical analysis of model~\eqref{improvedHJ} with $h_2=0$ and show that in case of $q>0$ one stable solution exists and in case of $q<0$ two stable solutions with different amplitudes may exist and the polarity of the solution depends on coefficient $p$.

The effect of the additional dispersive term $h_2u_{xxtt}$ on the numerical and analytical solutions has been analysed in previous studies \citep{Tamm2015,Peets2015,Peets2016,Engelbrecht2017}. The mixed derivative term does not only limit the speed of the higher harmonics \citep{Engelbrecht2015} but also controls the width of the solitary wave solution \citep{Peets2015,Peets2016,Engelbrecht2017} which is an effect related to the properties of the structure (inertia) of the lipid bilayer. It has been shown by numerical and analytical analysis \citep{Tamm2015,Peets2015,Engelbrecht2017} that solitary and oscillatory wave solutions may emerge with different sets of parameters demonstrating a rich spectrum of possible solutions emerging from Eq.~\eqref{improvedHJ}.

\section{Constant profile solutions }

For the convenience of further analysis Eq.~\eqref{improvedHJ} is written in a dimensionless form \citep{Peets2016}:
\begin{equation}
	\label{DimensiomlessImprovedHJ}
	U_{TT}=\left[(1+PU+QU^2)U_X\right]_X-H_1U_{XXXX}+H_2U_{XXTT},
\end{equation}
with $U=u/\rho_A$, $X=x/l$, $T=c_0t/l$, $P=p\rho_A/c_0^2$, $Q=q\rho_A^2/c_0^2$,\break $H_1=h_1/(c_0^2l^2)$ and $H_2=h_2/l^2$. Here $l$ is a certain length, for example, the diameter of the axon.

Solution that propagates at constant dimensionless velocity $c$ while preserving its shape can be written as $V=V(\xi)$ where $V$ is some function and $\xi=X-c T$ is a moving frame \citep{Drazin1989,Ablowitz2011,Dauxois2006}. Substituting this ansatz into Eq.~\eqref{DimensiomlessImprovedHJ} and integrating twice, a second order ODE \citep{Peets2016,Engelbrecht2017}:
\begin{equation}
	\label{iHJ2ODE}
	(H_1-H_2c^2)V''=(1-c^2)V+\frac{1}{2}PV^2+\frac{1}{3}QV^3+A\xi+B
\end{equation}
is  obtained, where $A$, $B$ are constants of integration and $(\;)'=d/d\xi$. This equation can be integrated with standard ODE solvers for numerical solutions. Further we will focus on solitary wave solutions and require that $V,V',V''\rightarrow 0$ as $X\rightarrow\pm\infty$ and therefore $A,B=0$ \citep{Ablowitz2011,Drazin1989,Dauxois2006}. 

For deriving analytic solutions, Eq.~\eqref{iHJ2ODE} is multiplied by $V'$ and integrated once to get 
\begin{equation}
	\label{PseudoPotPoly}
	(H_1-H_2c^2)(V')^2=(1-c^2)V^2+\frac{1}{3}PV^3+\frac{1}{6}QV^4.	
\end{equation}
Using an useful analogy form mechanics, Eq.~\eqref{PseudoPotPoly} can be interpreted as conservation of energy with the lhs acting as a kinetic term and the rhs as a `pseudo-potential' $\Phi_{eff}(V)$ \citep{Drazin1989,Ablowitz2011,Dauxois2006}. 

Next, Eq.\eqref{PseudoPotPoly} is rewritten as
\begin{equation}
	(V')^2=V^2\left[(V-a_+)(V-a_-)\right]\frac{Q}{6(H_1-H_2c^2)},
\end{equation}
where $a_\pm$ are the roots of a quadratic equation 
\begin{equation}
	\label{quadraticEq}
	QV^2/6+PV/3+(1-c^2)=0
\end{equation}
and a change of variable $V=1/y$ is used:
	\begin{equation}
		\label{ODEChangeOfVar}
		\left(-\frac{1}{y^2}\;y'\right)^2=\frac{1}{y^2}\left(\frac{1}{y}-a_+\right)\left(\frac{1}{y}-a_-\right)\frac{Q}{6(H_1-H_2c^2)}. 
	\end{equation} 
Then Eq.~\eqref{ODEChangeOfVar} is multiplied by $y^4$ and a series of straightforward algebraic operations are carried out, arriving to
\begin{equation}
	y'=\pm\sqrt{\frac{Qa_+a_-}{6(H_1-H_2c^2)}}\sqrt{(y-a)^2-b^2},
\end{equation}
where $a=(a_++a_-)/2a_+a_-$ and $b=(a_+-a_-)/2a_+a_-$ have been introduced for convenience. Since $y'=dy/d\xi$, then after integration we get
\begin{equation}
	\sqrt{\frac{Qa_+a_-}{6(H_1-H_2c^2)}}\xi=\pm\arccosh\left(\frac{y-a}{b}\right).
\end{equation}
Solving this equation for $y$ and then using $V=1/y$, the following solutions are obtained:
\begin{subequations}
	\label{HJsoliton}
	\begin{align}
	\label{HJsoliton1}
	&U_1(\xi)=\frac{-6(1-c^2)}{P+P\sqrt{1-6(1-c^2)Q/P^2}\cosh (\xi\sqrt{(1-c^2)/(H_1-H_2c^2)}) },\\
	\label{HJsoliton2}
	&U_2(\xi)=\frac{-6(1-c^2)}{P-P\sqrt{1-6(1-c^2)Q/P^2}\cosh (\xi\sqrt{(1-c^2)/(H_1-H_2c^2)}) }	.	
	\end{align}
\end{subequations}
\begin{figure}
\centering
\includegraphics[width=0.32\textwidth]{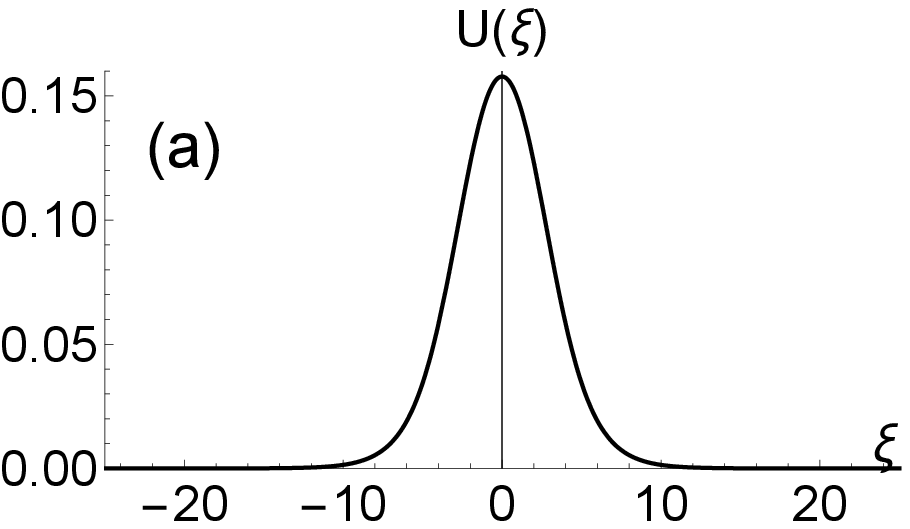}
\includegraphics[width=0.32\textwidth]{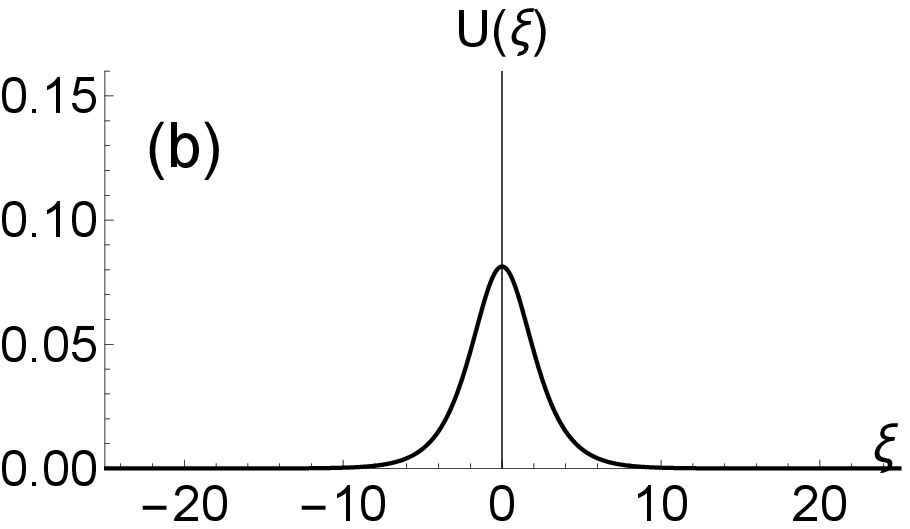}
\includegraphics[width=0.32\textwidth]{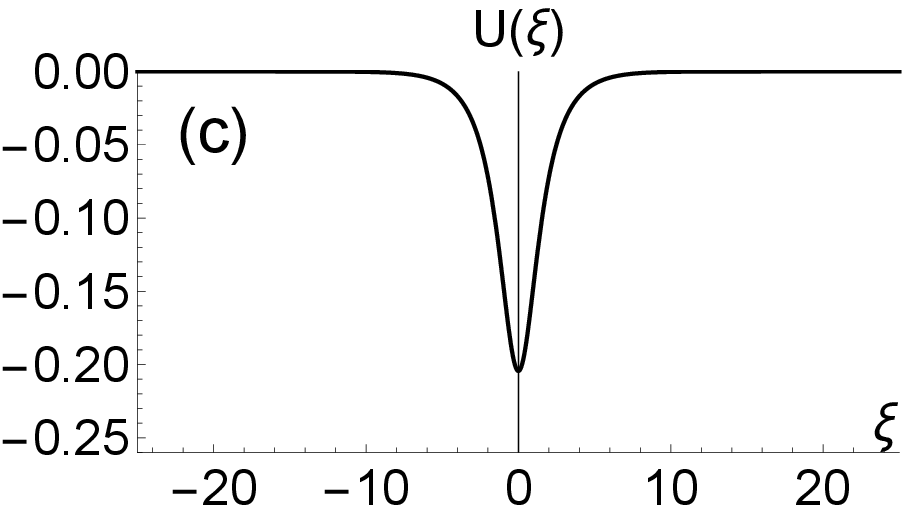}
\caption{Solutions of Eq.~\eqref{improvedHJ} in case of (a) $P=-10$, $Q=40$ and (b,c) $P=-8$, $Q=-130$. In all figures $c=0.8$, $H_1=4$ and $H_2=5$. In case of $Q>0$ only solution~\eqref{HJsoliton1} exists (a) and in case of $Q<0$ solutions~\eqref{HJsoliton1} and \eqref{HJsoliton2} coexist (b,c).  }
\label{Fig1}
\end{figure}

For solitary wave solutions the root inside hyperbolic cosine in Eq.~\eqref{HJsoliton1} has to be real (periodic solutions arise in case of imaginary root \cite{Engelbrecht2017}) and it is required that $(1-c^2)Q/P^2<1/6$, which sets limits to velocity $c$ \citep{Heimburg2005,Engelbrecht2017}. When aforementioned conditions are fulfilled and $(1-c^2)/(H_1-H_2c^2)>0$ then solution~\eqref{HJsoliton1} will always represent solitary wave solution and solution~\eqref{HJsoliton2} will only represent solitary wave solution in case of $Q<0$. This is demonstrated in Fig.~\ref{Fig1} and analysed in Section~4.

\section{Existence of soliton solutions }  

The solutions derived in previous Section are exact, but their interpretation needs some comments.
In this Section, the existence of solitary wave solutions is analysed with graphical analysis, which is ``clear and simple'' \citep{Strogatz1994}. For simplicity the further analysis is restricted for the case of $H_1-H_2c^2>0$. The results for the case of $H_1-H_2c^2<0$ are similar with some exceptions (see \citep{Engelbrecht2017} for details).

The classical approach for analysis of the existence of solitary waves is based on the polynomial on the rhs of Eq.~\eqref{PseudoPotPoly}, which is also known as the `pseudo-potential' \citep{Drazin1989,Ablowitz2011,Dauxois2006}:
\begin{equation}
	\label{EffPot}
	\Phi_{eff}(V)=(1-c^2)V^2+\frac{1}{3}PV^3+\frac{1}{6}QV^4.
\end{equation}
Solitary wave solutions exist when there is a double zero, zeros are real, \break $\Phi_{eff}(V)>0$ and there is a local maximum next to the double zero. 

\begin{figure}
\centering
\includegraphics[width=0.45\textwidth]{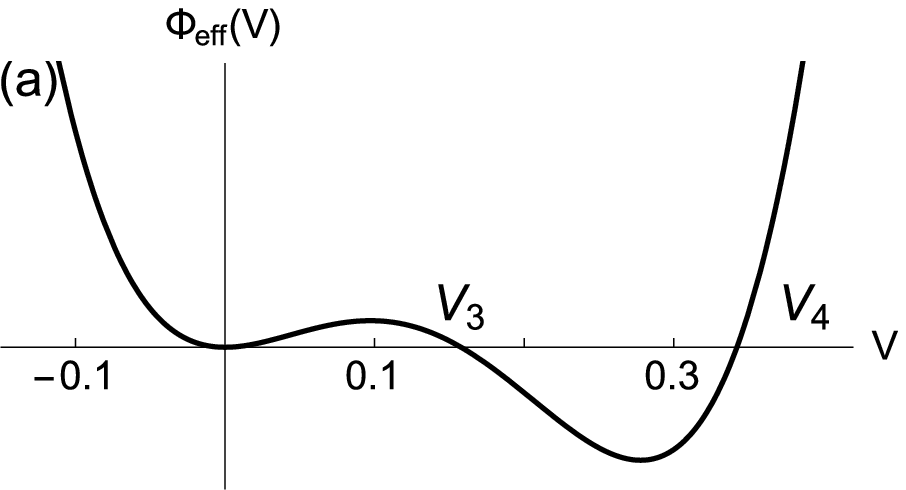}\quad
\includegraphics[width=0.45\textwidth]{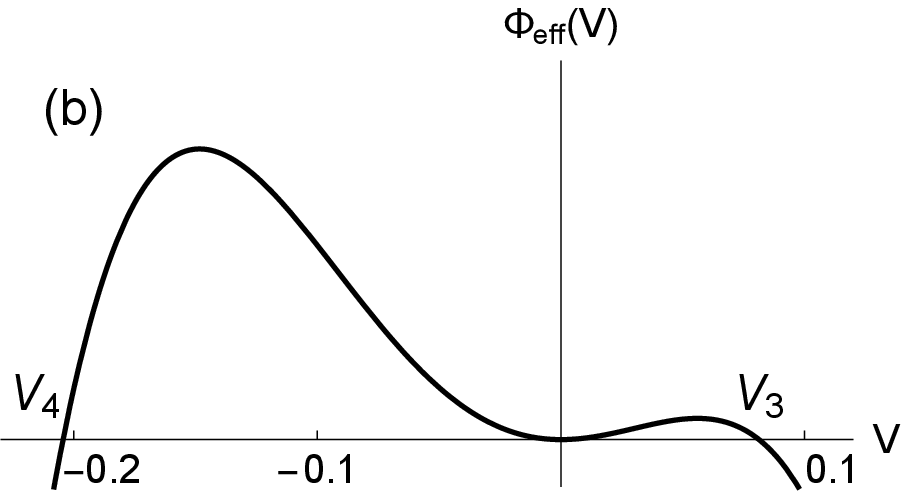}
\caption{Shape of the `pseudo-potential'~\eqref{EffPot} in case of $Q>0$ (a) and in case of $Q<0$. Coefficients are same as in Fig.~\ref{Fig1}: (a) $Q=40$; (b) $Q=-130$.}
\label{Fig2}
\end{figure}

For Eq.~\eqref{DimensiomlessImprovedHJ} the four zeros of the polynomial \eqref{EffPot} are 
\begin{equation}
	\label{zeros}
  V_{1,2}=0\quad \text{and}\quad V_{3,4}=P/Q\left(-1\pm\sqrt{1-6(1-c^2)Q/P^2}\right).
\end{equation}
Since the value of $V_{1,2}$ does not depend on the choice of coefficients, then the requirement for a double zero is always fulfilled at the origin. The values of zeros $V_3$ and $V_4$ depend on the choice of parameters -- in case of $Q>0$ the zeros of the polynomial \eqref{EffPot} have the same sign and there will be only one bounded region where $\Phi_{eff}(V)>0$. In case of $Q<0$ the zeros $V_3$ and $V_4$ have opposite signs and consequently two regions with $\Phi_{eff}(V)>0$ exist. This is demonstrated in Fig.~\ref{Fig2} for the case of $P<0$. For the case of $P>0$ the zeros $V_3$ and $V_4$ will have opposite signs and the shape of the `pseudo-potential' is flipped with respect to the vertical axis \citep{Peets2015,Engelbrecht2017}.

The dynamic behaviour of complicated nonlinear ODEs can be understood through the analysis of phase portraits. In addition to the existence of solitary wave solutions, this method gives insight to the existence of other kind of solutions. To that end Eq.~\eqref{iHJ2ODE} is rewritten as a system of first order ODEs:
	\begin{subequations}
		     \label{ODEsystem}
	\begin{align}
		&V'=W,\\
        &W'=(H_1-H_2c^2)^{-1}\left[(1-c^2)V+\frac{1}{2}PV^2+\frac{1}{3}QV^3\right].
		\end{align}
		\end{subequations}
Fixed points $V^\ast$ are found by setting $V'=W'=0$:
\begin{equation}
	\label{fixed points}
	V^\ast_1=0, \quad V^\ast_{2,3}=\frac{3}{4}\frac{P}{Q}\left(-1\pm
	\sqrt{1-\frac{16}{3}(1-c^2)Q/P^2}\right)
\end{equation}
and the nature of the fixed points is found by finding the eigenvalues   $\lambda$ of the Jacobian matrix for system~\eqref{ODEsystem} for each fixed point $V^\ast$ \citep{Strogatz1994}:
\begin{equation}
	\label{eigenvalues}
	\lambda_{1,2}=\pm i \sqrt{\frac{-PV^-QV^{2}-(1-c^2)}{H_1-H_2c^2}}.
\end{equation}
It can be seen in Eqs~\eqref{fixed points} and \eqref{eigenvalues} that while the coordinates of the fixed points depend on the nonlinear coefficients $P$, $Q$ and velocity $c$, the nature of the fixed points depends on the same coefficients and the dispersion type. In case of Eq.~\eqref{improvedHJ} the fixed points are either a saddle (real eigenvalues) or a centre (imaginary eigenvalues) depending on the choice of coefficients \citep{Engelbrecht2017}.
For the existence of solitary wave solutions there has to be a saddle point at the double zero and a homoclinic orbit. 

\begin{figure}
\centering
\includegraphics[width=0.45\textwidth]{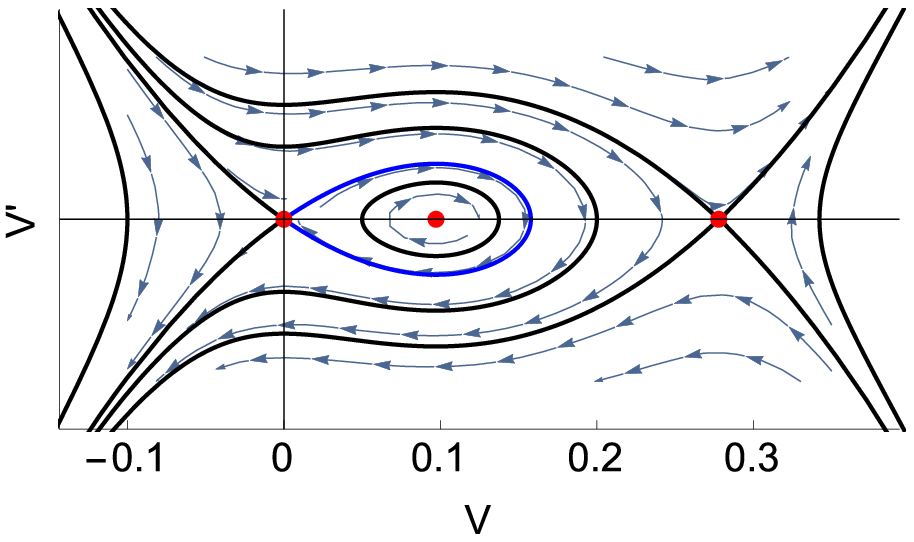}\quad
\includegraphics[width=0.45\textwidth]{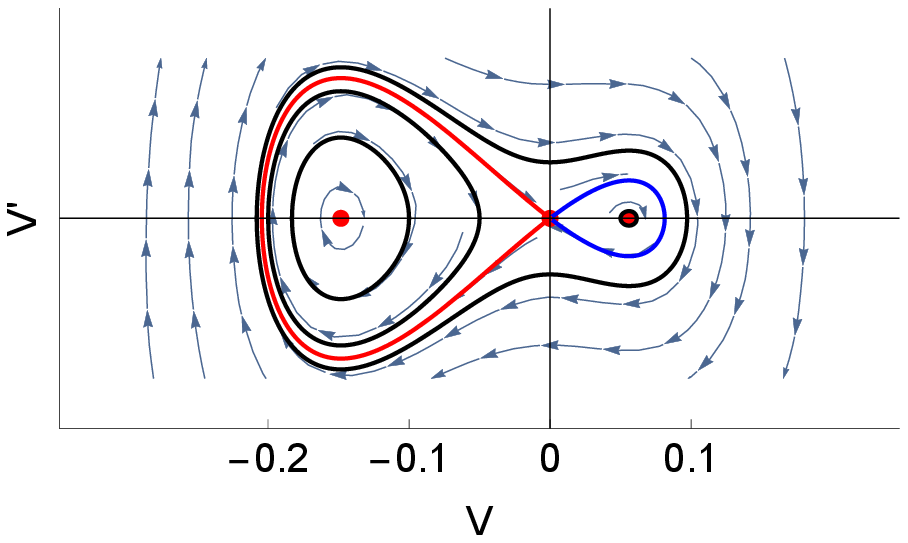}
\caption{Phase portraits in case of $Q>0$ (left) with $P=-10$, $Q=40$ and for the case of $Q<0$ (right) with $c=0.8$, $P=-8$, $Q=-130$. In both figures $c=0.8$, $H_1=4$ and $H_2=5$.}
\label{Fig3}
\end{figure}

In case of $H_1-H_2c^2>0$ and $c<1$ the fixed point $V^\ast_1$ is always a saddle and $V^\ast_2$ is always a centre. The nature of the fixed point $V^\ast_3$ depends on coefficient $Q$ -- it is a centre when $Q<0$ and a saddle when $Q>0$ \citep{Engelbrecht2017}. This is demonstrated in Fig.~\ref{Fig3} where the phase portrait for the case of $P<0$ are shown. It can be seen that in case of $Q>0$ only one homoclinic orbit exists and for the case of $Q<0$ two homoclinic orbits with different amplitudes exist. The homoclinic orbit representing solution \eqref{HJsoliton1} corresponds to the right loop (blue online) and the homoclinic orbit representing solution \eqref{HJsoliton2} corresponds to the left loop (red online) in Fig.~\ref{Fig3}, the fixed points are marked by solid dots (red dots online). For the case of $P>0$ the phase portraits are topologically similar, only the polarity of the homoclinic orbits are flipped. 

The analysis of the `pseudo-potential' (Fig.~\ref{Fig2}) and the phase portraits (Fig.~\ref{Fig3}) therefore indicate that in case of $Q<0$ Eq.~\eqref{improvedHJ} permits two solitary wave solutions with opposite polarity. In case of $Q>0$ only one solitary wave solution \eqref{HJsoliton1} exists. Mathematically also the solution~\eqref{HJsoliton2} exists but it does not represent a traveling wave.  In addition, it can be seen in the phase portraits that periodic solutions exist around centre points.

\begin{figure}
	\centering
	\includegraphics[width=0.45\textwidth]{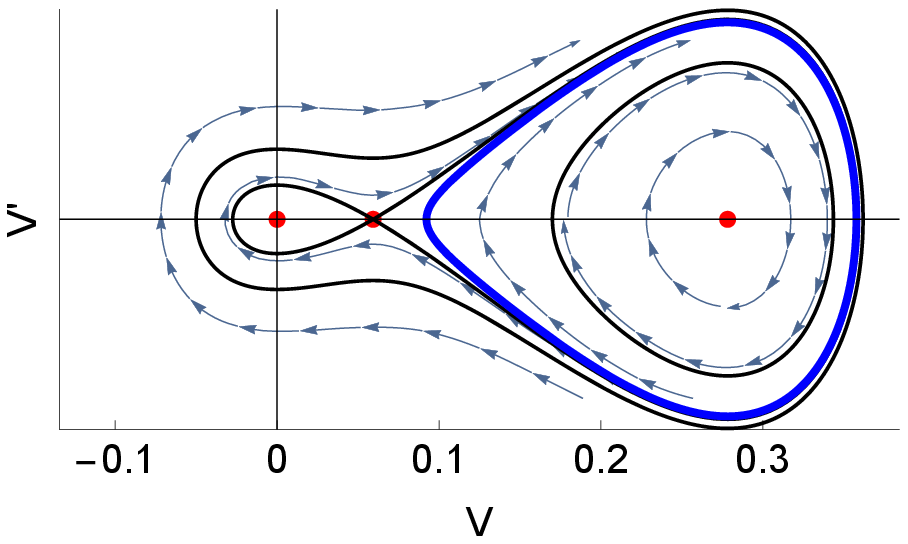}\quad
	\includegraphics[width=0.45\textwidth]{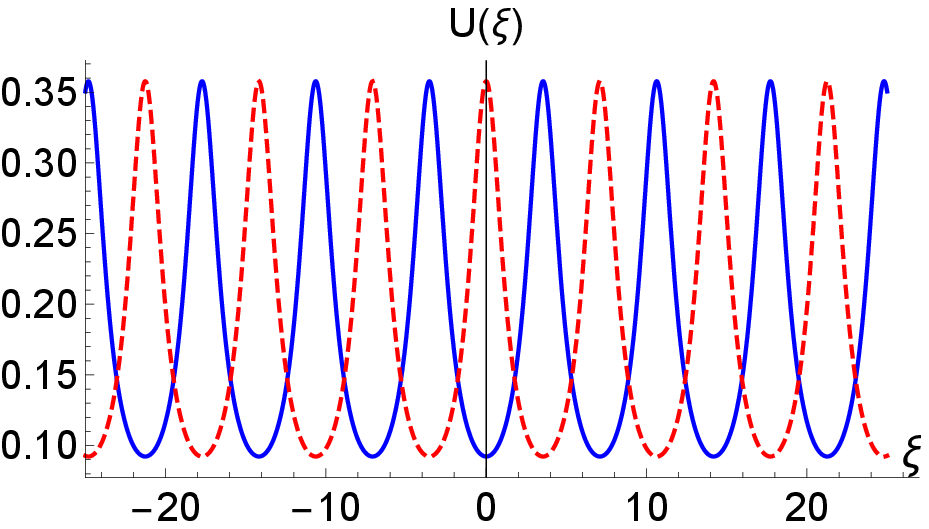}
	\caption{Solution~\eqref{HJsoliton} may also represent two periodic waves with a phase difference of $\pi$ when $(1-c^2)/(H_1-H_2c^2)<0$. Here left -- phase portrait, right -- wave profiles. Solution~\eqref{HJsoliton1} is shown in solid blue and solution~\eqref{HJsoliton2} in dashed red.  Here $c=1.2$, $P=18$, $Q=-80$, $H_1=2$ and $H_2=1$.}
	\label{periodic}
\end{figure}

\begin{figure}
	\centering
	\includegraphics[width=0.45\textwidth]{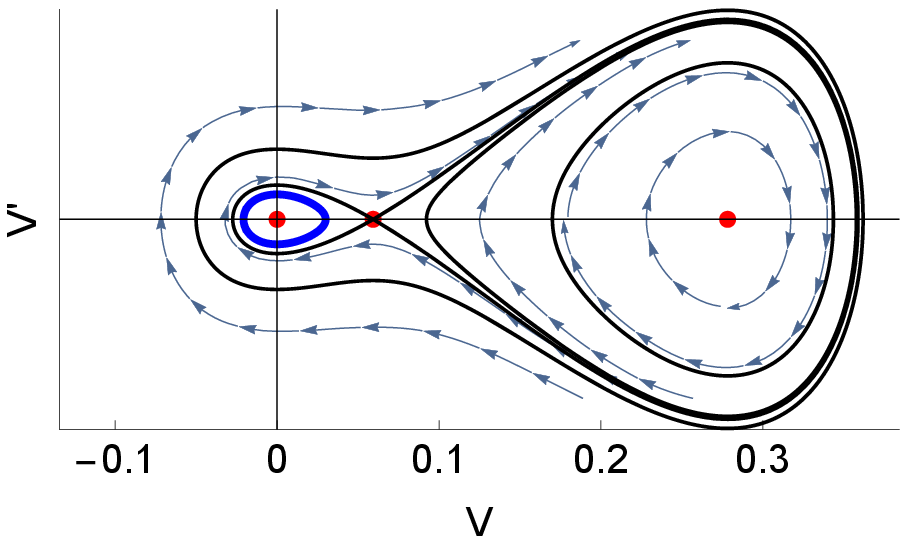}\quad
	\includegraphics[width=0.45\textwidth]{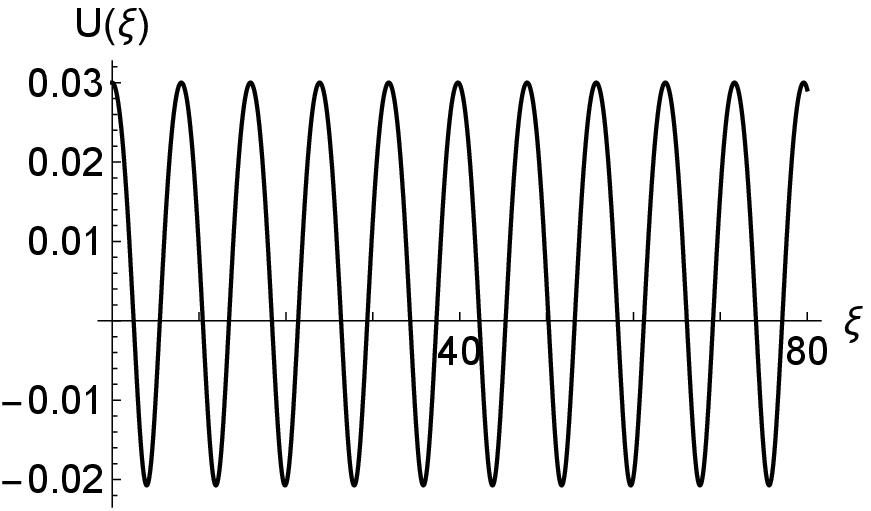}
	\caption{Fixed point at the origin ($V^\ast_1$) becomes a centre in case of $(1-c^2)/(H_1-H_2c^2)<0$ and periodic solutions around it can be obtained numerically. Here left -- phase portrait, right -- wave profiles. Coefficients are same as in Fig.~\ref{periodic}.}
	\label{FigLeftLoop}
\end{figure}

As shown in the previous analysis \citep{Engelbrecht2017}, the exact solutions~\eqref{HJsoliton} also represent periodic solutions when $(1-c^2)/(H_1-H_2c^2)<0$ which means that the fixed point $V^\ast_1$ becomes a centre and solitary wave solutions are no longer possible. Transition from solitary wave solutions to periodic solution can also be seen in solution~\eqref{HJsoliton} since the imaginary root under hyperbolic cosine follows the aforementioned condition. In this case solutions~\eqref{HJsoliton1} and \eqref{HJsoliton2} represent identical coexisting periodic waves with a phase difference of $\pi$. This is demonstrated in Fig.~\ref{periodic} where the periodic waves are within the right loop of the homoclinic orbit. Note that the parameters for the phase portrait in Fig.~\ref{periodic} differ from those in Fig.~\ref{Fig3} satisfying the condition presented above. It is known \citep{Ablowitz2011} that analytical solitary wave solutions exist when the double zero \eqref{zeros} is also a saddle point. Although in case of $(1-c^2)/(H_1-H_2c^2)<0$  the double zero remains at the origin, the fixed point at double zero is a centre (see Fig.~\ref{periodic}). The periodic waves within the left loop of the phase portrait are obtained numerically and shown in Fig.~\ref{FigLeftLoop}.

\section{Soliton doublets}

\begin{figure}
	\centering
	\includegraphics[width=0.8\textwidth]{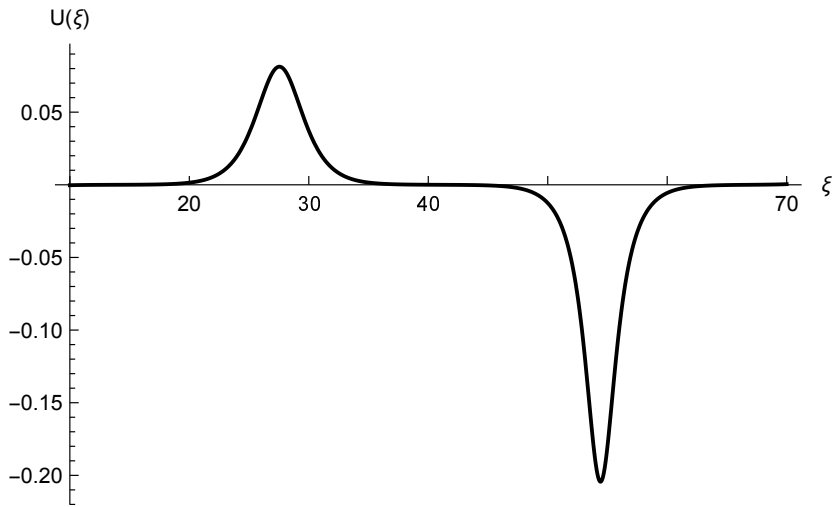}
	\caption{The numerical solution of Eq.~\eqref{improvedHJ} demonstrating the soliton doublet corresponding to the phase space depicted in Fig.~\ref{Fig3} (right). Here $c=0.8$, $P=-8$, $Q=-130$, $H_1=4$ and $H_2=5$.}
	\label{Fig4}
\end{figure}

It was shown in previous Section that if $(1-c^2)/(H_1-H_2c^2)<0$, then in case of $Q<0$ Eq.~\eqref{improvedHJ} has two solitary wave (or periodic) solutions. Question that arises is whether two solitary wave solutions can coexist. We will first show that there exists a trajectory in the phase space that represents a soliton doublet (coexisting solitary waves with same velocities but different amplitudes). To that end the system~\eqref{ODEsystem} is integrated by an ODE solver \cite{WolframResearch2018} with a suitable initial amplitude -- either $V_3$ or $V_4$ (see Eq.~\eqref{zeros}). We seek for a solution which is slightly out of the homoclinic orbit. The solution demonstrating `nearly solitary' waves \citep{Randruut2014} which may be called a `doublet soliton', is shown in Fig.~\ref{Fig4}. The parameters in Fig.~\ref{Fig4} are same as in Fig.~\ref{Fig3} where the homoclinic orbits responsible for solitary wave solutions are shown in the phase portrait. Note that in calculations, the doublet will be repeated after long time. The co-existing solitons are described by expressions~\eqref{HJsoliton1} and \eqref{HJsoliton2} where the free parameters $c$, $P$ and $Q$ determine the amplitude of solitons (cf. with the classical soliton \citep{Ablowitz2011}).

\begin{figure}
	\centering
	\includegraphics[width=0.8\textwidth]{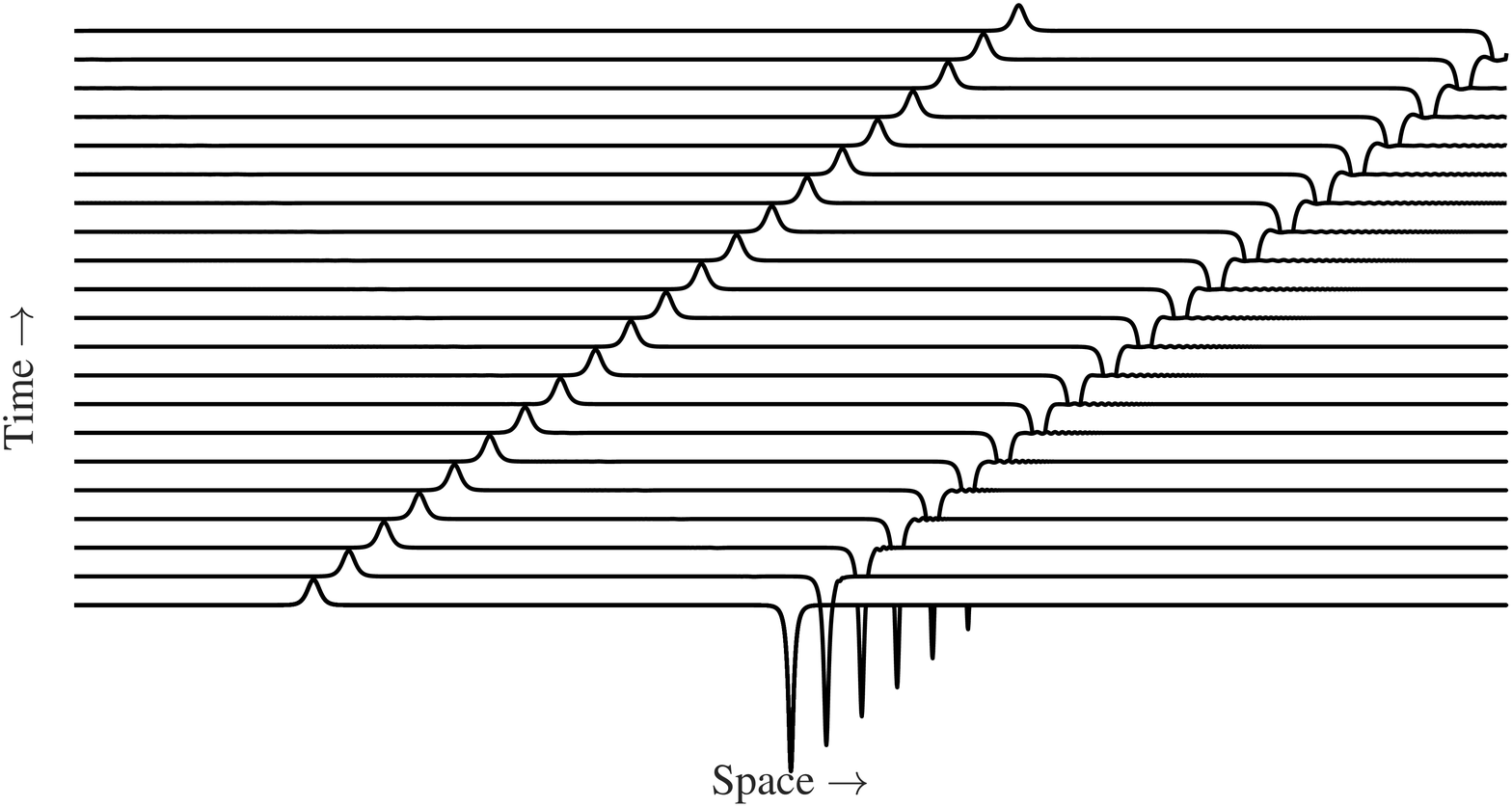}
	\caption{A soliton doublet: time slice plot. Here $c=0.99$, $P=-0.5$, $Q=-1.7946877$, $H_1= 0.5$, $H_2=0.25$. The total length of the space is $512\pi$ and a segment from $128\pi$ to $512\pi$ is shown and $\Delta T=30$.}
	\label{Fig6}
\end{figure}

We will next show that a soliton doublet will propagate at constant velocity while maintaining its shape. The solution is obtained numerically by making use of the Discrete Fourier Transform (DFT) based pseudospectral method (PSM) \citep{Salupere2009,Engelbrecht2017}. Using this method, the governing equation needs to be in a specific form with only time derivatives in the right-hand side and only spatial derivatives in the left-hand side of the equation. In order to deal with the mixed derivative term in Eq.~\eqref{improvedHJ} a new variable $\Phi$ is introduced and the variable $U$ and its spatial derivatives are expressed in terms of this new variable 
\begin{equation}
	\label{spekraalmeetod}
	\Phi = U - H_2 U_{XX},\, U=\mathrm{F}^{-1}\left[\frac{\mathrm{F}(\Phi)}{1+H_2 k^2}\right], 
\,
\frac{\partial^m U}{\partial X^m}
=\mathrm{F}^{-1}\left[\frac{(\mathrm{i} k)^m \mathrm{F}(\Phi)}{1+H_2 k^2}\right],
\end{equation}
where $\mathrm{F}$ denotes the Fourier transform and $\mathrm{F}^{-1}$ is the inverse Fourier transform; $k$ is discrete frequency. Equation~\eqref{improvedHJ} in its dimensionless form \eqref{DimensiomlessImprovedHJ} is then rewritten as 
\begin{equation}
	\Phi_{TT} =
 (1+PU+QU^2)U_{XX}
+(P+2QU)\left(U_X\right)^2-H_1U_{XXXX}. 
\end{equation}
This equation can be easily solved by making use of the PSM after reducing it to a system of two first-order differential equations (see~\citep{Salupere2009} for details). The sum of solutions represented by Eq.~\eqref{HJsoliton1} and spatially shifted Eq.~\eqref{HJsoliton2} is used as an initial pulse with corresponding time derivatives as an initial condition.

The result of the numerical simulation is shown in Fig.~\ref{Fig6}, where it is seen that a soliton doublet propagates while retaining the separation of the constituents (distance between the peaks of the pulses is $128\pi$). This means that for a given set of parameters, Eq.~\eqref{improvedHJ} permits the existence of a soliton doublet, i.e., a solution where two nearly solitary waves with different amplitudes propagate at the same velocity.

\section{Discussion}

Equation~\eqref{improvedHJ} (or its dimensionless form~\eqref{DimensiomlessImprovedHJ}) describes longitudinal waves in a special nonlinear medium -- a biomembrane. The nonlinearity is of the displacement-type and dispersive terms account for the elasticity ($u_{xxxx}$) and the inertia ($u_{xxtt}$) of the microstructure (lipid molecules). The soliton-type solutions of Eq.~\eqref{improvedHJ} have been described earlier  \citep{Heimburg2005,Engelbrecht2015}. The full mathematical analysis of Eq.~\eqref{improvedHJ} is carried out \citep{Engelbrecht2017} including also the analysis of the emergence of soliton trains and of the processes of interaction of solitons. The analytical solution of Eq.~\eqref{HJsoliton1} is known \citep{Lautrup2011} for the case of $h_2=0$ and is generalised for the case of $h_2\neq 0$ \citep{Peets2016}. Here it has been shown that Eq.~\eqref{improvedHJ} has one more solution which in case of $Q<0$ presents either a solitary or periodic wave. The phase portrait (Fig.~\ref{Fig3}, right) demonstrates clearly the existence of two homoclinic orbits and it is seen that the `pseudopotential' \eqref{EffPot} has two regions where $\Phi_{eff}(V)>0$. The corresponding solutions \eqref{HJsoliton1} and \eqref{HJsoliton2} present different solitons. In addition, this paper turns attention to a special case of possible solutions of Eq.~\eqref{improvedHJ}: the existence of a soliton doublet.

The existence of soliton doublets is described already by Scott et al.~\citep{Scott1973} and in principle, N-soliton solutions are known from the general analysis by inverse scattering transform \citep{Ablowitz2011}. The forced Korteweg-deVries (KdV) system \citep{Engelbrecht2005a} as well as the hierarchical KdV-type systems \citep{Ilison2009} may lead to solitonic structures or soliton ensembles. Recently the doublets are described in laser systems \citep{Grelu2006}, in Heisenberg ferromagnetic model \citep{Gerdjikov2011}, fibre Bragg gratings \citep{Chen2008}, etc. It seems that like in the present case, the nonlinearities play a decisive role in forming the doublets (cf. non-Kerr nonlinearity in the nonlinear Schr\"odinger equation, described by Azzouzi et al. \citep{Azzouzi2015}). The notion of `a dipole' for describing soliton doublets is used by Hermon et al. \citep{Hermon1998} for waves in DNA molecules and by Min et al. \citep{Min2016} for waves in metamaterials. In both these cases complicated nonlinearities play also a crucial role in emerging the special solitary complexes.

\section*{Acknowledgements}
This research was supported by the European Union through the European Regional Development Fund (Estonian
Programme TK 124) and by the Estonian Research Council (projects IUT 33-24, PUT 434). 

\section*{References}

\end{document}